\documentclass[aps,prl,twocolumn, amsmath,amssymb,showpacs,superscriptaddress]{revtex4-2}  % for review and submission
\usepackage{graphicx}  % needed for figures
\usepackage{dcolumn}   % needed for some tables

\usepackage{bm}        % for math
\usepackage{amsmath}
\usepackage{amsfonts}
\usepackage{amssymb}%
\usepackage{mathrsfs}

\usepackage{mathabx}

\providecommand{\U}[1]{\protect\rule{.1in}{.1in}}

\include{settings}

\begin{document}
\title{Gravitational Effect on Fundamental Physical Observables}
\author{Yoshimasa Kurihara}
\email{yoshimasa.kurihara@kek.jp}
\affiliation{High Energy Accelerator Research Organization, Tsukuba, Ibaraki 305-0801, Japan}
\date{\today}
\begin{abstract}
This report studies possible gravitational effects on measurements of fundamental physical observables such as the fine structure constant and the lepton magnetic moment.
Although a static gravitational potential does not contribute to local observables in the inertial system owing to Einstein's equivalence principle, a dynamic degree with finite momentum transfer can contribute to them.
A laboratory frame fixed on the Earth's surface is not in an inertial system; thus, Earth's gravity can contribute to local observables.
We investigate experimental results measuring the fine structure constant and electron and muon magnetic moments using the quantum field theoretic method under the Schwarzschild background field.
We prepare classical vierbein and spin-connection fields, owing to the Schwarzschild metric in the momentum space of the local Lorentz space and give estimations of possible gravitational effects at the surface of the Earth.
The gravitational effect does not contribute to the fine structure constant measurements.
On the other hand, the Earth's gravity gives an $\OO(10^{-9})$ effect on the muon anomalous magnetic moment that is consistent with the measured values. 
 
We propose a possible test of gravitational effects in particle physics using precise measurement of muon properties in the J-PARK. 
\end{abstract}
%03.65.Sq Semiclassical theories and applications
%13.40.Em Electric and magnetic moments
%13.40.Gp Electromagnetic form factors
%13.87.Ce Production
%04.60.−m Quantum gravity
%04.60.Bc Phenomenology of quantum gravity
%04.62.+v Quantum fields in curved spacetime
%04.70.Dy Quantum aspects of black holes, evaporation, thermodynamics
%\keywords{test}
\maketitle
The standard theory of elementary particles\cite{10.1093/ptep/ptac097} describes the microscopic aspects of the Universe very accurately, except for a few observations.
One example of an experimental observation showing a contradiction in a theoretical prediction is the precise measurement of the muon anomalous magnetic moment, which deviates from theoretical predictions by about three standard deviations\cite{Muong-2:2006rrc,Muong-2:2021ojo}.
On the other hand, that of an electron shows reasonable agreement\cite{PhysRevLett.100.120801}.
We refer to it as the $\gm2$ anomaly in this report.
Measured and SM expected values of the anomalous magnetic moment of both an electron and a muon are summarized in TABLE-\ref{g-2}, together with the fine structure constant measurements.
\begin{table*}[bt]
 \caption{\small The anomalous magnetic moment and the fine structure constant for an electron and a muon.
 Numbers in parentheses are values of an indirect measurement obtained from theoretical calculations.
 Errors in ``exp.-theor.'' is a linear sum (not a quadratic sum) of errors for the experiment and the theoretical calculation.}
% \centering
  \begin{tabular}{llccc}
  ~&exp./theor. &$(g-2)/2$&$\alpha^{-1}$& Ref. \\    
    \hline\hline\vspace{-2.5mm}\\
  electron  &exp.($g$/2)  &$1159 652 180590(\hspace{0.5em}13)\times10^{-14}$&$\(137035999166(15)\times10^{-9}\)$&\cite{Fan:2022eto}\\
~                 &exp.(Cs)  &$\(1159 652 181610(\hspace{0.5em}23)\times10^{-14}\)$&$137035999046(27)\times10^{-9}$&\cite{Parker_2018}\\
~                 &exp.(Rb)  &$\(1159 652 180252 (\hspace{0.5em}95)\times10^{-14}\)$&$137035999206 (11)\times10^{-9}$&\cite{Morel:2020dww}\\
~                 &theor.&$1159 652 181606(253)\times10^{-14}$&$137035999150(36)\times10^{-9}$&\cite{Aoyama:2019ryr}\\
~&exp.($g$/2)-theor.&$-9.1(2.6)\times10^{-13}$&\\\vspace{-3mm}\\
  \hline\vspace{-2.5mm}\\
 muon  &exp.  &$116592061(41)\times10^{-11}$&~&\cite{Muong-2:2006rrc,Muong-2:2021ojo}\\
~           &theor.&$116 591 810(43)\times10^{-11}$&(using a value of exp.(Cs)\cite{Parker_2018})&\cite{Aoyama:2020ynm}\\
~&exp.-theor.  &$+2.5(0.8)\times10^{-9}$&\\
  \hline
  \end{tabular}
 \label{g-2}
\end{table*}

Morishima, Futamase and Shimizu\cite{Morishima_2018} proposed a possible resolution to the $\gm2$ anomaly owing to the static gravitational potential of the Earth using the post-Newtonian approximation.
Visser\cite{Visser:2018omi} immediately criticised their proposal as contradicting Einstein's equivalence principle and pointed out also that they missed counting the effect of the Sun's and the Galaxy's static potential, which are much stronger than the Earth's.
Following Visser's criticism, Nikoli\'{c}\cite{Nikolic:2018dap} and Venhoek\cite{Venhoek:2018biz} clarified a flaw in calculations by Morishima et al.

Discussions mentioned above are based on classical electromagnetism and general relativity and do not treat the dynamical aspect of gravity in common.
A frame fixed on the Earth is the inertial system concerning the Sun's and the Galaxy's gravity.
Due to Einstein's equivalence principle, the static gravitational potential does not contribute to any local observables obtained in the inertial system.
On the other hand, experimental apparatus on the Earth is not in an inertial system but in an accelerating system concerning the Earth's gravity.
Finite momentum is transferred to the system to keep it fixed on the Earth.

This report treats the Earth's gravitational effect from a dynamic point of view.
Using the quantum field theoretic method in the Schwarzschild background field, we propose an alternative resolution of the $\gm2$ anomaly owing to the Earth's gravity.
Although we use Planck units setting $c=\hbar=\kappa=1$ in this study, we write $\hbar$ and $\kappa$ explicitly.

%\vskip 2mm
Suppose $(\MM,\bm{g})$ is a four-dimensional (pseudo) Riemannian manifold, where $\MM$ is a smooth and oriented four-dimensional manifold, and $\bm{g}$ is a metric tensor with a negative signature in $\MM$.
In an open neighbourhood $U_{\hspace{-.1em}p\hspace{.1em}{\in}\MM}\subset\MM$, we introduce the standard coordinate $x^\mu$.
The standard orthonormal bases in $\TMM_{p}$ and $\TsMM_{p}$ are denoted as $\partial/\partial x^\mu$ and $dx^\mu$, respectively.
We use the abbreviation $\partial_\mu:=\partial/\partial x^\mu$ throughout this report.
Two trivial vector bundles $\TMM:=\bigcup_{p} \TMM_{p}$ and $\TsMM:=\bigcup_{p} \TsMM_{p}$ are referred to as a tangent and cotangent bundles in $\MM$, respectively.

An inertial system, in which the Levi-Civita connection vanishes, exists locally at any point in $\MM$.
An inertial frame at point $p\in\MM$ denoted as $\M_{p}$ is referred to as a local inertial manifold at point $p$.
Suppose $\M_{p}$ has a $SO(1,3)$ symmetry.
We denote the standard coordinate as $\xi^a$ in an open neighbourhood $U_{\hspace{-.1em}p\hspace{.1em}{\in}\M}\subset\M$.
Trivial bundles $\TM:=\bigcup_{p} T\M_{p}$ and $\TsM:=\bigcup_{p} \TsM_{p}$ also exist over $\M$.
As for suffixes of vectors in $\TM_{p}$, Roman letters are used for components of the standard basis; Greek letters are used for them in $\MM_{p}$.
Owing to the convention, we can distinguish two abbreviated vectors such that $\partial_\mu\in V(\TMM)$ and $\partial_a=\partial~/\partial\xi^a\in V(\TM)$.
The metric tensor in $\M$ is $\bm{\eta}=\textup{diag}(1,-1,-1,-1)$.
The Levi-Civita tensor (complete anti-symmetric tensor) $\bm\epsilon$, whose component is $[\bm\epsilon]_{0123}=\epsilon_{0123}=+1$, and $\bm\eta$ are constant tensors in $\TsM$.

We define the vierbein $\E^a_\mu(x)\in{C^\infty(\MM)}$ as a map transferring a vector in $\TsMM$ to that in $\TsM$ such that:
\begin{align*}
\E^a_\mu(x\in\TsMM)dx^\mu\big|_{p\hspace{.1em}\in\MM_p}=d\xi^a\big|_{p\hspace{.1em}\in\M_p}
\in\Omega^1(\TsM){\otimes}\sss\ooo(1,3),
\end{align*}
where $\Omega^p(\bullet)$ denotes a space of $p$-form objects defined in manifold $\bullet$.
The vierbein is a smooth and invertible function globally defined in $\MM$.
The vierbein inverse $[\E^{-1}]_a^\mu=\E_a^\mu\in{C^\infty(\M)}$, which is also called the vierbein, is an inverse transformation such that
\begin{align*}
\E^a_\mu(x)\E_a^\nu\(\xi(x)\)&=\delta^\nu_\mu\in\TsMM,\\
%\intertext{and}
\E_a^\mu\(\xi\)\E^b_\mu(x\(\xi)\)&=\delta_a^b\in\TsM.
\end{align*}
Covariant differential $d_\www$ concerning an $SO(1,3)$ group for one-form object $\aaa^a\in\TsM$ is defined using spin connection $\omega^{~ac}_{\mu}$ as
\begin{align*}
d_\www\aaa^a&:=d\aaa^a+\cG\hspace{.1em}\www^a_{~\bcdot}\wedge\aaa^\bcdot,
\end{align*}
where
\begin{align*}
\www^a_{~b}&:=\omega^{~a\bcdot}_{\mu}\hspace{.1em}\eta_{\bcdot b}\hspace{.1em}dx^\mu
\in \Omega^2(\TM)\otimes\Omega^1(\TsMM)\otimes\sss\ooo(1,3).
\end{align*}
We introduce the gravitational coupling constant $\cG$, which does not appear standardly in general relativity.
The coupling constant can be absorbed by simultaneously scaling a connection and a curvature when a single structure group is in the principal bundle. 
In this study, coupling constants provide a relative strength of couplings for two structure (gauge) groups, $SO(1,3)$ and $SU(N)$.

Raising and lowering indices are done using a metric tensor.
Dummy Roman indices are often abbreviated to a small circle $\bcdot$ (or $\star$) when the dummy-index pair of the Einstein convention is obvious.
When multiple circles appear in an expression, the pairing must be in a left-to-right order at upper and lower indices.
The spin connection is an $SO(1,3)$ Lie-algebra valued anti-symmetric tensor in $\TsM$ (concerning upper Roman indices) and a vector in $\TsMM$ (concerning lower Greek index) belongs to the adjoint representation of an $SO(1,3)$  group.
Detailed definitions and geometric structure of space-time and local inertial manifolds are summarized in Refs.\cite{Kurihara:2022sso,Kurihara:2022green}. 

In the Yang--Mills theory with the $SU(N)$ gauge group, the Lagrangian density of a massive charged fermion is given as\cite{Kurihara:2022sso}
\begin{align*}
\LL_{\psi}
&=
\bar{\bm{\psi}}\left(i\hspace{.1em}\gamma^\bcdot\partial_\bcdot-i\frac{\cG}{2}
\gamma^\bcdot\E_\bcdot^\mu\omega_\mu^{~\stars}S_\stars
+%\right.\notag\\&\hspace{5.5em}+\left.
{c^{~}_\SU}\hspace{.1em}\gamma^\bcdot\Aa^I_\bcdot \tau^{~}_I-\frac{m}{\hbar}
\right)\bm{\psi}%\label{LagYM}
\end{align*}
where $\tau^I$ is a Lie-algebra of the gauge group, ${c^{~}_\SU}$ is a gauge coupling constant, and 
$S_{ab}:=i\sigma_{ab}/2=i[\gamma_a,\gamma_b]/4$ is a generator of the $Spin(1,3)$ group.
For simplicity, we write no $SU(N)$ unit matrix in the Lagrangian.
Spinors $\bm{\psi}:=(\psi_1,\cdots,\psi_n)^T$ exists only in the $SO(n)$ space\cite{PhysRev.101.1597,RevModPhys.36.463}; thus, spinor field $\psi_i$ and Clifford algebra $\gamma^a$ are defined in $\TsM$. 
Gauge field $\Aa^I_a$ is a connection concerning the structure group of the gauge bundle defined in the local inertial manifold; thus, it is defined in $\TsM$.
On the other hand, spin connection $\omega_\mu^{~ab}$ is a function defined in $\TsMM$.
After \emph{Romanization} of the lower index, the spin connection is a function in $\TsM$ such that:
\begin{align*}
\omega^{~bc}_a(\xi\in\TsM)&=\E_a^\mu\hspace{.1em}\omega_\mu^{~bc}\in
C^\infty(\TsM).
\end{align*}
Consequently, the Yang--Mills Lagrangian density is a smooth function defined in $\TsM$.

A classical Yang--Mills field strength is a curvature owing to the $SU(N)$ gauge connection in the local inertial manifold given as\cite{Kurihara:2022sso}
\begin{align*}
\f^I_{ab}&:=\partial_a\Aa^I_b-\partial_b\Aa^I_a+
{{c^{~}_\SU}}\hspace{.1em}f^I_{~JK}\Aa^J_a\Aa^K_b+\Aa^I_\bcdot\TT^\bcdot_{~ab},
\end{align*}
where $f^I_{~JK}$ is a structure constant of $SU(N)$ and $\TT$ is a torsion tensor of the inertial manifold.
When the space-time manifold is torsionless, as in Einstein's gravity, the gauge curvature in the curved space-time has the same representation as that of the flat space-time.
Hereafter, we treat a $U(1)$ gauge group in the torsionlass space, and set $\bm{\psi}\rightarrow\psi$, $\Aa^I_\mu\rightarrow\Aa_\mu$, $\tau^I\rightarrow1$, ${c^{~}_\SU}\rightarrow e$, $f^I_{~JK}\rightarrow0$, and $\TT\rightarrow0$ in the formulae given above.

The classical vierbein fields of the Schwarzschild metric using polar-coordinate $dx^\mu=(dt,dr,d\vartheta,d\varphi)$ in $\TsMM$ are
\begin{subequations}
\begin{align}
\E^{a=0}_{0}dx^0&=\fsch(r)\hspace{.1em}dt,&
\E^{a=1}_{1}dx^1&=\fsch(r)^{-1}\hspace{.1em}dr,\label{sch1}\\
\E^{a=2}_{2}dx^2&=r\hspace{.1em}d\vartheta,&
\E^{a=3}_{3}dx^3&=r\sinvt\hspace{.1em}d\varphi,\label{sch2}
\end{align}
\end{subequations}
otherwise zero, where $\fsch(r):=\sqrt{1-\rs/r}$ and $\rs$ is the Schwarzschild radius.
% and $M$ is a mass of a gravity source.
The vierbein provides a determinant as det$[\bm\E]=r^2\sin{\vartheta}$.

The Fourier transformation of the field in the configuration space provides that in the momentum space. We denote a space of vectors Fourier-transformed from those in $\TsM$ as $\TstM$.
The current author discussed the Fourier transformation and the momentum space in curved space-time in detail\cite{Kurihara:2022green}.
In reality, the classical vierbein fields of the Schwarzschild metric in the momentum space using polar-coordinate $dq^a=(dE,dq,d\vartheta,d\varphi)$ in $\TstM$ with the non-zero three-momentum $|\vec{q}|>0$ are obtained from (\ref{sch1}) and (\ref{sch2}) such that\cite{Kurihara:2022sso}:
\begin{subequations}
\begin{widetext}
\begin{align}
\tilde{\E}_{0}\(\pb\)=
\tilde{\E}^{1}\(\pb\)
&\simeq-\frac{{\pi\hspace{.1em}\rs^3}}{2}
\(\frac{1}{\pb^{2}}+\frac{\pi}{4\pb}-\frac{1}{2}\log{\frac{\pb}{2}}+
\frac{1}{12}\(1-6\gamma^{~}_E-6i\pi\)
+{\cal O}\(\pb^{1}\)\),\label{Eb}\\
\tilde{\E}_{1}\(\pb\)=
\tilde{\E}^{0}\(\pb\)
&\simeq\frac{{\pi\hspace{.1em}\rs^3}}{2}
\(\frac{1}{\pb^{2}}+\frac{3\pi}{4\pb}-\frac{5}{2}\log{\frac{\pb}{2}}-\frac{1}{12}\(30\gamma^{~}_E+30i\pi+7\)
+{\cal O}\(\pb^{1}\)\),\label{Ed}
\end{align}
\end{widetext}
\end{subequations}
where $\pb:={\rs}\hspace{.1em}|\vec{q}\hspace{.08em}|/(2\hbar)$ is a dimension-less momentum variable.
In (\ref{Eb}) and (\ref{Ed}), we use notations
\begin{align*}
\tilde{\E}_{i}\(\pb\):=\tilde{\E}_{\mu=i}^{a=i}\(\pb\) ,&~~
\tilde{\E}^{i}\(\pb\):=\tilde{\E}^{\mu=i}_{a=i}\(\pb\),
\end{align*}
for simplicity.
For the asymptotic observer at $r\rightarrow\infty\Leftrightarrow\pb\rightarrow0$, we require a normalisation of the vierbein field as
\begin{align*}
\tilde{\E}_{i}\(\pb\)\xrightarrow{\pb\rightarrow0}\frac{1}{\pb^2}.
\end{align*}
Consequently, we obtain a normalisation factor
\begin{align}
f^{~}_{\hspace{-.1em}\E}:=\(2\pi\hspace{.1em}\rs^3\)^{-1},\label{Enorm}
\end{align}
yielding
\begin{align*}
\pb^2\left|\bar{\E}_{E}\(\pb\)\right|&:=\pb^2f_{\hspace{-.1em}\E}\hspace{.1em}\tilde{\E}_{0}\(\pb\)
\simeq1+\frac{\pi}{4}\pb,\\
\pb^2\left|\bar{\E}_{z}\(\pb\)\right|&:=\pb^2f_{\hspace{-.1em}\E}\hspace{.1em}\tilde{\E}_1\(\pb\)
\simeq1+\frac{3\pi}{4}\pb,
\end{align*}
around $\pb\simeq0$ .

The momentum transfer against the Earth's gravity is typically $\pb^{~}_\Earth{\simeq}\kappa M^{~}_\Earth/r^{~}_\Earth{\simeq}6.95\times10^{-10}$, where $r^{~}_\Earth$ and $M^{~}_\Earth$ are Earth's radius and mass, respectively.
Therefore, we expect a possible effect from the Earth's gravity, such as 
\begin{align*}
\pb^2_\Earth\left|\bar{\E}_{E}\(\pb^{~}_\Earth\)\right|-1&\simeq5.46\times10^{-9},\\
\pb^2_\Earth\left|\bar{\E}_{z}\(\pb^{~}_\Earth\)\right|-1&\simeq1.64\times10^{-9}.
\end{align*}
The Earth is a freely falling object concerning gravity from the Sun, the Galaxy and other gravitational sources; thus, the momentum transfer from objects fixed on the Earth to those other than the Earth is zero.
We note that the estimated gravitational effect is comparable to the sensitivity of precise measurements of the fine structure constant and the lepton magnetic moment.

First, we look at precise measurements of the fine structure constant, whose measured value is used to estimate the theoretical prediction of the lepton magnetic moment.
Two independent methods have measured the fine structure constant; one deduces it from a measured $m_e/m_a$ value and the Rydberg constant, and the other calculates it  from a measured value of the electron magnetic moment, where $m_a$ is a mass of the atom used for experiments, i.e., Cs or Rb.
The $2S$-$12D$ two-photon transitions on atomic hydrogen and deuterium provide the Rydberg constant\cite{PhysRevLett.82.4960}.
In \cite{PhysRevLett.82.4960}, laser photons interact with free-falling atoms in a vacuum of the inertial system; thus, no gravitational effect occurs in this measurement.
Also, $m_e/m_a$ measurements were performed in the inertial system with removing the gravitational effect carefully\cite{Parker_2018,Morel:2020dww}.
In conclusion, no gravitational effect is in the fine structure constant measurements using the atomic system.
The electron magnetic moment measurement also provides indirect information on the fine structure constant independently from the atomic method mentioned above. 
Possible gravitational effects on the electron magnetic moment measurement are discussed later in this report.

Next, we discuss the muon magnetic moment measurement:
Experimental apparatuses are set on the Earth, which is fixed in the global manifold $\MM$; thus, the magnetic field inducing a precession of a lepton spin is defined in $\TsMM$.
We set the gauge connection in $\TsMM$ as $\Aa_{\mu}=(A_0,A_x,A_y,A_z)$, where we take a Cartesian coordinate such that the $x$-axis along to a beam momentum, the $y$-axis to the centre of the accelerator ring and the $z$-axis perpendicular to the Earth's surface.
This report refers to it as the lab-frame. 
The gauge connection in the local inertial manifold is 
\begin{align*}
\Aa_a(\xi)=\E_a^\mu\Aa_\mu=(\fsch(r^{~}_\Earth)^{-1}A_0,A_x,A_y,\fsch(r^{~}_\Earth)A_z).
\end{align*}
The BNL--FNAL type magnetic moment measurement of a muon\cite{Muong-2:2006rrc,Muong-2:2021ojo} applies the magnetic field along the $z$-axis, which is provided by the gauge connection of $\Aa_{\mu}=\Aa_a=(0,-yB,xB,0)/2$, where $B$ is a magnetic field strength of analysing magnets. 
Consequently, the gauge connection $\Aa_a(\xi)$ has no gravitational term, and the Earth's gravity does not affect the muon magnetic moment measurement.

The electron magnetic moment is measured for a single electron stored in the Penning trap with a strong magnetic field\cite{PhysRevA.47.2610,PhysRevLett.100.120801}.
The analysing magnetic field is also along the $z$-axis; thus, the gravitational effect does not occur owing to the same reason as the BNL--FNAL type experiments.
As a result, the fine structure constant obtained from the electron magnetic moment are also free from the Earth's gravity.

So far, we have discussed the possible gravitational effect owing to the third term of the Lagrangian.
Although the second term does not couple to the $U(1)$ connection, it induces an additional precession to the rotating spin.

A three-dimensional spacial Fourier transformation of $\omega_c^{~ab}(\xi)$ provides the spin connection in the momentum space using polar-coordinate $dq^a$ such that\cite{Kurihara:2022sso}:
\begin{widetext}
\begin{align*}
{\tilde{\omega}^{~10}_0(\pb)}
&\simeq\frac{\pi}{2}\rs^2\(\frac{\pi}{\pb}-2\log{\frac{\pb}{2}}
-2\gamma^{~}_E-2i\pi+{\cal O}\(\pb\)\),\\
%%%
{\tilde{\omega}^{~12}_2(\pb)}=
{\tilde{\omega}^{~13}_3(\pb)}=
&\simeq
\frac{\pi}{2}\rs^2\(
\frac{2}{\pb^2}-\frac{\pi}{\pb}
+\log{\frac{\pb}{2}}+\gamma^{~}_E-\frac{1}{2}-i\pi
+{\cal O}\(\pb\)\),
\end{align*}
\end{widetext}
and otherwise zero.
We set the $x$-axis to a beam momentum $p^{~}_e$ in the lab-frame for experiments using an electron beam; thus,  on-shell beam momenta are\cite{Kurihara:2023caf}
\begin{align*}
p^{~}_{\text{lab}}&=\(E^{~}_{e},p^{~}_{e},0,-\frac{q^{~}_e}{2}\) ,~~~~
p'_{\text{lab}}=\(E'_{e},p^{~}_{e},0,\frac{q^{~}_e}{2}\),
\end{align*}
yielding $p^{2}_{\text{lab}}=p'^{2}_{\text{lab}}=m^2_e$, and $q^{~}_\text{lab}:=p'_{\text{lab}}-p^{~}_{\text{lab}}$, where $q^{~}_e$ is momentum transfer between an electron and the Earth through the spin-connection.
The Schwarzschild spin-connection in the Cartesian coordinate is
\begin{align*}
{\tilde\omega}_{0}^{\hspace{.2em}03}(\pbL)&=-{\tilde\omega}_{0}^{\hspace{.2em}30}(\pbL)\\
&=\frac{\pi}{2}\(\frac{\Rs}{\hbar}\)^2
\(\frac{\pi}{\pbL}-2\log{\frac{\pbL}{2}}-2\gamma^{~}_E+{\cal O}\(\pbL\)\),\\
{\tilde\omega}_{1}^{\hspace{.2em}13}(\pbL)&=-{\tilde\omega}_{1}^{\hspace{.2em}31}\\
&=\frac{\pi}{2}\(\frac{\Rs}{\hbar}\)^2
\(\frac{2}{\pbL}+\pbL\log{\frac{\pbL}{2}}-\pi+{\cal O}\(\pbL\)\),
\end{align*}
otherwise zero.
We set them to have an inverse energy-squared dimension.

To estimate the gravitomagnetic effect quantitatively for the $\cG$ measurement, we utilise the Breit frame in which we have $p^{~}_{\text{Br}}+p'_{\text{Br}}=(2m^{~}_e,0,0,0)$ with neglecting $\OO(q^{~}_e/m^{~}_e)$, such that\cite{Kurihara:2023caf}:
\begin{align*}
p^{~}_{\text{Br}}&=
\(m^{~}_e-\frac{\sqrt{E^{2}_e-m^2_e}}{2m^{~}_2}q^{~}_e,\hspace{.6em}\frac{E^2_e-m^2_e}{2E^{~}_em^{~}_e}q^{~}_e,0,-\frac{q^{~}_e}{2}\) ,\\
p'^{~}_{\text{Br}}&=
\(m^{~}_e+\frac{\sqrt{E^{2}_e-m^2_e}}{2m^{~}_2}q^{~}_e,-\frac{E^2_e-m^2_e}{2E^{~}_em^{~}_e}q^{~}_e,0,\hspace{.5em}\frac{q^{~}_e}{2}\) ,
\end{align*}
ans $q^{~}_\text{Br}=p'_{\text{Br}}-p^{~}_{\text{Br}}$.
We ignore effects due to the centrifugal and Coriolis forces in following calculations.
In the Breit frame, the spin-connection of the Schwarzschild solution in the momentum space is obtained fomr those in the lab-frame after the Lorentz boost.  
%In the following, we discuss the scattering amplitude in the Breit frame and omit the subscript ``Br'' for simplicity.

We give the scattering amplitude as
%\begin{subequations}
\begin{align}
i{\tau}^{\text{S}}_\Earth=
\bar{u}^{\lambda'}\hspace{-.2em}\hspace{-.2em}(p')\hspace{.1em}V^{\textrm{S}}\hspace{.1em}u^{\lambda}_{~}\hspace{-.1em}(p),\label{tauS}
\end{align}
with
\begin{align*}
V^{\textrm{S}}:=-i\frac{\cG}{2}
\tilde{\omega}^{\text{S}\hspace{.2em}\stars}_{\hspace{.4em}\bcdot}\gamma^\bcdot\frac{\sigma_\stars}{2},
\end{align*}
owing to the Lagrangian.
%\end{subequations}
We insert an identity\footnote{
When $p$ is an on-shell momentum, ${\eta_{*s}^{~}\gamma^* p^*+m^{~}_e}$ does not have an inverse.
We consider $p$ to be slightly off-shell and perform the following calculations.
In the end, we put it back to unity.
}
\begin{align*}
1&=\frac{\sum_{\lambda''}u^{\lambda''}\hspace{-.2em}(p)\hspace{.2em}\bar{u}^{\lambda''}\hspace{-.2em}(p)}
{\eta_{**}^{~}\gamma^* p^*+m^{~}_e}
\end{align*}
into (\ref{tauS}), yielding
\begin{align*}
\text{(\ref{tauS})}&=-i\frac{\cG}{2}
\tilde{\omega}^{\text{S}\hspace{.2em}\stars}_{\hspace{.4em}\bcdot}
\sum_{\lambda''}\bar{u}^{\lambda'}\hspace{-.2em}(p')\gamma^\bcdot
\frac{u^{\lambda''}\hspace{-.2em}(p)\hspace{.2em}\bar{u}^{\lambda''}\hspace{-.2em}(p)}
{\eta_{**}^{~}\gamma^* p^*+m^{~}_e}
\frac{\sigma_\stars}{2}{u}^\lambda(p).%\label{Vwpsi2}
\end{align*}
We look at an electron current in the above formula and use the Gordon identity, yielding
\begin{align}
&\bar{u}^{\lambda'}(p')\gamma^au^{\lambda''}\hspace{-.2em}(p)\notag\\&=
\bar{u}^{\lambda'}(p')\(\frac{p^a+p'^a}{2m^{~}_e}+
\frac{i\sigma^{a\bcdot} (p-p')^\bcdot\eta_\bcdots}{2m^{~}_e}
\)u^{\lambda''}\hspace{-.2em}(p).\label{Gordon}
\end{align}
The second term is in $\mathcal{O}(q^{~}_e/m^{~}_e)$ and negligible with respect to the first term.
Thus, we have
\begin{align*}
\text{(\ref{Gordon})}&\simeq\frac{1}{2m^{~}_e}\bar{u}^{\lambda'}(p')\(p^a+p'^a\)u^{\lambda''}\hspace{-.2em}(p)
\simeq
\bar{u}^{\lambda'}(p')\hspace{.1em}u^{\lambda''}\hspace{-.2em}(p)\hat{p}^{a}_{~},
\end{align*}
where $p^a+p'^a\simeq2m^{~}_e\hat{p}^{a}_{~}$ and $\hat{p}^{a}_{~}$ is a unit vector parallel to $p^a$.
We put this relation back into the interaction vertex and rearrange order of a polarisation sum again, yielding that
\begin{align*}
i{\tau}^{\text{S}}_\Earth&\simeq-i\cG\hspace{.1em}
\tilde{\bm{\omega}}^{~}_{\bcdot}\cdot\bm{S}\hspace{.1em}\hat{p}^{\bcdot}_{~},
\intertext{where}
\tilde{\bm{\omega}}^{a}_{~}\cdot\bm{S}&:=\frac{1}{2}\eta^{a\star}
\tilde{\omega}^{\hspace{.2em}\bcdots}_{\star}S_\bcdots,
\intertext{with}
S_{ab}&:=\eta^{~}_{a\bcdot}\eta^{~}_{b\bcdot}\bar{u}^{\lambda'}\hspace{-.2em}(p')\frac{\sigma_{~}^\bcdots}{2}u^{\lambda}(p).
\end{align*}
Consequently, we obtain the amplitude with the forward scattering approximation that
\begin{align*}
i{\tau}^{\text{S}}_\Earth&=
i\cG\(\tilde{{\omega}}^{0}_{~}(\pb)\hspace{.1em}S_{30}
-\tilde{{\omega}}^{1}_{~}(\pb)\hspace{.1em}S_{31}
-\tilde{{\omega}}^{2}_{~}(\pb)\hspace{.1em}S_{32}
\)\hat{p}^{3}_{~}.
\end{align*}
with
\begin{align*}
S^{~}_{30}&\simeq\hspace{.8em}\bm\xi'^\dagger\(p^{~}_x\sigma^2-p^{~}_y\sigma^1\)\bm\xi,\\
S^{~}_{31}&\simeq\hspace{.8em}{m^{~}_e}\hspace{.1em}\bm\xi'^\dagger\sigma^2\bm\xi,\\
S^{~}_{32}&\simeq-{m^{~}_e}\hspace{.1em}\bm\xi'^\dagger\sigma^1\bm\xi.
\end{align*}
Here, we write a $y$-component even it is zero.

We define a gravitomagnetic-moment vector of an electron as
\begin{align*}
\bmuG:=\frac{\cG}{m^{~}_e}\gG\hspace{.1em}\vec{s}
~~~\text{with}~ ~~\vec{s}=\hspace{.2em}\bm\xi'^\dagger\frac{\vec{\sigma}}{2}\hspace{.1em}\bm\xi,
\end{align*}
where $\vec{s}$ is a spin angular-momentum vector and $\gG$ is a \emph{gravitomagnetic $g$-factor}.
A gravitomagnetic energy of an electron in a gravitational field $\bmBG$ is 
\begin{align*}
\VG(r)&=-\bmuG\cdot\bmBG=-\frac{\cG}{m^{~}_e}\gG\hspace{.1em}\vec{s}\cdot\bmBG.%\label{Hgr}
\end{align*}

Amplitude ${\tau}^{\text{S}}_\Earth$ provides the Born approximation of the scattering of an electron with the potential due to the Earth's gravity:
 \begin{align*}
 i\bm\tau^{~}_{\text{Born}}=-i\(2m^{~}_e\)\bm\xi'^\dagger\tilde{V}(\pb)\bm\xi=i{\tau}^{\text{S}}_\Earth;
\end{align*}
thus, we obtain that
\begin{align*}
\tVG(0)&=-\frac{\cG}{m^{~}_e}\gG\hspace{.1em}{s^y_{~}}\BG~~\text{with}~~
\gG=2,
\end{align*}
in the Breit frame.
A coefficient of an angular momentum operator $s^y_{~}=\bm\xi'^\dagger(\sigma^2/2)\hspace{.1em}\bm\xi$ gives a rotation angle of an electron spin; thus, amplitude $i{\tau}^{\text{S}}_\Earth$ generates a counter-clockwise spin rotation around a $y$-axis such that an initial horizontal (along a $x$-axis) spin vector to a $z$-axis (upward).

When $\pb\neq0$, we can obtain the gravitational $g$-factor as
\begin{align*}
\gG(\pb)&=-\(\frac{\cG}{m^{~}_e}{s^y_{~}}\BG\)^{-1}\tVG(\pb),
\end{align*}
and it induces that a counterclockwise spin rotation around a $y$-axis such that an initial horizontal (along a $x$-axis) spin vector to a negative $z$-axis (upward).
We obtain an anomalous gravitomagnetic moment as 
\begin{align}
\aG&:=
\frac{\gG-2}{2}=\frac{1}{2}\(\gG\(\pb\)-\gG\(0\)\).\label{gg-2}
\end{align}
The anomalous magnetic-moment measurements may include an effect of the above contribution.

In the electron anomalous magnetic-moment measurements\cite{PhysRevLett.100.120801,Fan:2022eto}, an electron has a small Lorentz factor $\gamma^{~}_e$ and $\gamma^{~}_e\beta^{~}_e$.
Moreover, these experiments utilise free-falling electrons with $\pb\simeq0$.
Thus, the spin precession owing to the gravitomagnetic effect is negligible compared with the magnetic one. 

On the other hand, the BNL--FNAL type $\gmt$ measurement used the ``magic momentum''\cite{Muong-2:2006rrc,Muong-2:2021ojo} of $p=4.094$GeV  corresponding to Lorentz factors $\gamma=29.4$ and $\beta=0.999421$ to eliminate the spin precession due to the focusing electric field; thus, we can expect a sizable precession owing to the gravitomagnetic moment with the Earth's gravity.
%The electrostatic quadrupole (ESQ) covers $13/30$ of the muon storage ring\cite{Muong-2:2015xgu}.
Muons in the storage ring are kept horizontally due to the electric field of ESQ on average. 
Therefore, we estimate each muon receives a momentum transfer of $\pb^{~}_\Earth$ on average, which induces the spin precession owing to (\ref{gg-2}), such that:
\begin{align*}
\cG\hspace{.1em}\aG&=257\hspace{.2em}858\times10^{-11}.
\end{align*}
A total precesstion effect is estimated as
\begin{align*}
a^{2}_\text{total}&:=\aSM^2+\(\cG\hspace{.1em}\aG\)^2+2\cG\hspace{.1em}\aSM\hspace{.1em}\aG\cos{\theta_a},
\end{align*}
where $\aSM$ is the theoretical prediction in Table \ref{g-2} and $\theta_a$ is a angle between two precession axes.
The precession owing to the anomalous grativomagnetic moment is around the $y$-axis and that owing to the anomalous magnetic moment is around the $z$-axis in the lab-frame.
Consequently, $\theta_a=\pi/2$ and the total anomalous magnetic moment, including the gravitomagnetic contribution, yields \begin{align*}
\Delta\aG&:=\(\left.a^{~}_\text{exp}-a^{~}_\text{total}\)\right|_{\cG=1}=-3.4\times10^{-10}.
\end{align*}
consistent with the current $\gm2$ measurement, when we set $\cG$ to unity.

An independent $\gm2$  measurement of the BNL--FNAL type experiment is under construction at J-PARC\cite{10.1093/ptep/ptz030}.
This experiment utilizes low-emittance muons confined in a magnetic field (along the $z$-axis in the beam frame) in a weak electric field.
Moreover, a $\beta$ factor is smaller than the BNL--FNAL type experiments by about one order of magnitude. 
Therefore, there is a negligible gravitomagnetic effect, and our analysis expects that they may obtain a consistent $\gm2$ value with the standard model prediction without the gravitational effect.

We can expect additional effects when a magnetic field is rotated horizontally.
When $x$- and $y$-axes are fixed horizontally, and the $z$-axis is perpendicular to the horizontal plane, the gauge connection in $\TsMM$ is $\Aa_{\mu}=(0,0,-zB,yB)/2$ and the synchrotron trajectory is in the $y$-$z$ plane.
The gauge connection in the local inertial manifold $\M$ is 
\begin{align*}
\Aa_a(\xi)=\frac{B}{2}(0,0,-z,y\fsch(r^{~}_\Earth)),
\end{align*}
and a magnetic field deffers by $B\fsch(r^{~}_\Earth)/2$  from that in $\MM$.
An expected effect on the $\gm2$ measurement is $\Delta_\E:=(\pb^{2}_\Earth\bar{\E}_{z}\(\pb^{~}_\Earth\)-1)/2$ in the momentum space of $\MM$, which provides
\begin{align*}
\Delta_\E\simeq8.2\times10^{-10}.
\end{align*}
This amount of the gravitational effect is comparable to the expected sensitivity of the J-PARC $\gm2$ experiment.
Although, a vertical circulation of the muon beam in BNL--FNAL type experiments is not realistic, an analyzing magnet of the J-PARC $\gm2$ experiment has a cylindrical shape with $3.56$m diameter and $3.73$m height and can presumably be rotated the magnet by 90 degrees from vertical to horizontal.

Our field theoretic analysis shows that a possible gravitational effect on the muon magnetic moment measurement owing to the Earth gravity is about $\OO(10^{-9})\sim\OO(10^{-10})$.
The J-PARC $\gm2$ experiments may provide a test bench to investigate the gravitational effects on particle physics.

\vskip 2mm
\begin{acknowledgments}
I would like to thank Dr Y$.$ Sugiyama, Prof$.$ J$.$ Fujimoto and Prof$.$ T$.$ Ueda for their continuous encouragement and fruitful discussions.
\end{acknowledgments}

% references
\bibliography{HRDg2}      % Bibliography file (usually '*.bib' )
\end{document}